\documentclass[journal]{IEEEtran}

\usepackage[hidelinks]{hyperref}
\usepackage{amsmath}
\usepackage{color}
\usepackage[font={small}]{caption}
\usepackage{amsfonts}
\usepackage{amssymb}
\usepackage{pifont}
\usepackage{epsf}
\usepackage{graphicx}
\usepackage{cite}
\usepackage{empheq}
\usepackage{acronym}
\usepackage{epstopdf}
\usepackage{comment}
\usepackage{float}
\usepackage{subcaption}
\usepackage{amsthm}
\usepackage{enumerate}   

\graphicspath{{./figs/}}
\DeclareGraphicsExtensions{.pdf}

\newcommand{\bfg}[1]{\boldsymbol{#1}}
\newcommand{\bfb}[1]{\boldsymbol{\rm #1}}

\newcommand{\unitvec}[1]{\bfb e_{#1}}

\newcommand{\Wgeom}[1]{\mathit{\hat{\Omega}}_{#1}}
\newcommand{\Div}[1]{\nabla \cdot \bfg #1}
\newcommand{\Trace}[1]{{\rm tr}(\bfb #1)}
\newcommand{\dt}{\partial_t}

\newcommand{\bivec}[1]{\hat{\bfb #1}}
\newcommand{\Wbiv}[1]{\bivec{\Omega}_#1}
\newcommand{\vort}{\mathbf{w}}

\begin{document}

\title{Equivalence between Geometric Frequency \\ and Lagrange Derivative}

\author{
  Federico~Milano,~\IEEEmembership{IEEE Fellow}%
  \thanks{F.~Milano is with the School of Elec.~\& Electron.~Eng., University College Dublin, Dublin, D04V1W8, Ireland.  
  e-mail: federico.milano@ucd.ie}%
  \thanks{This work is supported by the Sustainable Energy Authority of Ireland~(SEAI) under project FRESLIPS, Grant No.~RDD/00681.}%
  }

\maketitle

\begin{abstract}
  The paper shows the equivalence between the geometric frequency of an electric quantity, namely, voltage and current, and the Lagrange derivative of a stream-line of a fluid.  The geometric frequency is a concept recently proposed by the author and is a generalization of the instantaneous frequency, a quantity that is particularly important for the analysis and the control of electric power systems.  On the other hand, the Lagrange derivative is mostly utilized in fluid dynamics and helps decomposing the time derivative into various components.  The paper shows how these components relate to the elements of the geometric frequency.  The paper also shows, through a variety of numerical examples, how the decomposition of the Lagrange derivative helps identifying the distortion of the waveform of a measured electric quantity and how this information can be utilized to classify system operating conditions.
\end{abstract}


\IEEEpeerreviewmaketitle


\section{Introduction}
\label{sec:intro}

The estimation of the instantaneous frequency of electrical signals is important in power system analysis and operation \cite{FDF:2020}.  Main applications are the frequency control of inverter-based energy resources, over- and under-frequency protections, wide-area measurement systems and monitoring of the overall power balance of the system \cite{IEEEStd1459}.  However, the estimation of instantaneous frequency is a challenging task, especially in real-time applications and for signals in transient conditions or that include noise, harmonics and/or unbalances \cite{8586583, 8973973}.   

A variety of approaches have been proposed to the estimation of instantaneous frequency, for example, phase-locked loops~(PLLs) \cite{liu2014three, escobar2014cascade, kulkarni2015design, reza2019three, eskandari2022robust, sahoo2021phase}, discrete Fourier transform \cite{romano2013interpolateddft, song2022fast}, inverse Park transform \cite{santos2008comparison}, Hilbert transform \cite{hao2007measuring}, Kalman filters \cite{reza2016accurate, nie2019detection}, least squares, \cite{giarnetti2015non, pradhan2005freq}, and adaptive notch filters \cite{wilches2020method, mojiri2007estimation}.  The existence of such a variety of techniques suggests that no method proposed so far is clearly superior to the others and that the adequateness of each method, while certainly showing some advantages, depends on the application.

Reference \cite{freqgeom} has recently proposed the formal analogy between the voltage (current) at a point of an electrical multi-phase circuit and a velocity.  This analogy is supported by the Faraday's law, which states that the voltage is the time derivative of a magnetic flux and the definition of electric current as the time derivative of the flow of electric charges.  According to this analogy, the magnetic flux (electric charge) are assumed to be generalized coordinates.  This analogy has been further exploited in \cite{instapower} to link the geometric frequency and the energy stored in inductances and capacitances and the instantaneous active and reactive powers to the second derivative of the angular momentum commonly utilized in classical mechanics for the study of rigid bodies.

The geometric approach has the advantage to require only instantaneous measurements and is thus free of the issues such as aliasing and spectral leakage that affect Fourier-based methods.  On the other hand, since it requires to estimate the time derivative of the measured signal, its performance is comparable to methods based on PLLs, which are affected by noise, unbalances and harmonics.

In this work, we remove the hypothesis of rigid body and extend the analogy above to fluid mechanics.  In particular we assume that magnetic fluxes (electric charge flows) in every point of a multi-phase circuit represent \textit{stream lines} of the flow of a fluid.  It is important to note that, according to this assumption, each point of the circuit is a stream line, as the fluxes, not the spatial position, are the ``coordinates'' of the voltage.

The study of fluid mechanics is largely founded on the so-called Lagrange derivative, which shows the dependency of any time-dependent quantity on the velocity of the stream-line of the flow of a fluid \cite{Petrila:2005, Kambe:2007}.  In this work we are interested in the expression of the Lagrange derivative when it is applied to the velocity itself, which in our analogy is the voltage (current) at a point of the electric circuit.  When applied to the velocity, the Lagrange derivative can be split into various symmetric and skew-symmetric terms, each of them with very specific meaning, such as strain, distortion and rotation, of the element of the fluid stream line.

The main contribution of this work is the theoretical appraisal of the equivalence of the terms that compose the Lagrange derivative of the velocity with the terms that compose the geometric frequency of the voltage (current).  These terms provide a novel classification of the operating conditions of ac and dc circuits.  For example, the paper shows that a dc circuit can be studied as an irrotational flow, whereas a stationary ac circuit can be studied as an solenoidal flow.  Moreover, unbalances and harmonics can be interpreted in terms of the symmetric and skew-symmetric components of the Lagrange derivative.  Finally, as a byproduct, the paper shows the link between instantaneous frequency and the \textit{vorticity}, a quantity that is commonly utilized in fluid mechanics to describe the local spinning motion of a continuum \cite{Kambe:2007}.

The motivation for this work stems from the observation that the geometric frequency does not coincide, in many practical cases, with the quantity that in electric power systems, one would consider (or expect) to be the instantaneous frequency of the voltage.  For example, the geometric frequency is not constant for an unbalanced and/or non-sinusoidal system, even if the system is stationary and its fundamental frequency is constant.  This work shows that the vorticity is the quantity that actually matches the intuition of instantaneous frequency of an ac circuit.  

The remainder of the paper is organized as follows.  Section \ref{sec:frequency} recalls the definition of geometric frequency proposed in \cite{freqgeom} and shows that such a definition can be applied, in fact, to any time-dependent vector as it is, in turn, a way to express the time derivative of the vector itself.  Moreover, if the vector is a (generalized) velocity, such an expression can be interpreted in terms of geometric invariants.  Section \ref{sec:lagrange} recalls the Lagrange derivative and the various symmetric and skew-symmetric term into which this derivative can be decomposed.  Section \ref{sec:equivalence} provides the main contribution of this paper and shows the term-to-term equivalence of the geometric frequency and the Lagrange derivative.  Section \ref{sec:examples} illustrates the theoretical findings of Section \ref{sec:equivalence} though a variety of analytical examples.  Finally, Section \ref{sec:conclusions} draws conclusions and outlines future work.

\section*{Notation}

Unless otherwise indicated, scalars are represented in Italic, e.g., $x$, $X$; vectors in lower case bold, e.g., $\bfg x = (x_1, x_2, \dots, x_n)$; bivectors in upper case bold with a hat, e.g., $\hat{\bfg X}$; second order tensor and matrices in bold upper case, e.g., $\bfb X$, and multivectors in upper case Italic with a hat, e.g., $\hat{X}$.

\vspace{2mm}

\subsubsection*{Scalars}

\begin{itemize}[\IEEEiedlabeljustifyl \IEEEsetlabelwidth{Z} \labelsep 0.9cm]
\item[$h$] harmonic order
\item[$v_{\alpha}, v_{\beta}$] Clarke's components of the voltage %
\item[$\theta$] voltage phase angle %
\item[$\kappa$] magnitude shape factor for unbalanced voltage %
\item[$\xi$] frequency correction factor for unbalanced voltage %
\item[$\rho_t$] radial strain due to local time dependency %
\item[$\rho_r$] radial strain due to shear strain distortion %
\item[$\rho_v$] radial frequency of the voltage %
\item[$\tilde{\varphi}$] scalar potential for irrotational filed %
\item[$\omega_o$] fundamental frequency %
\end{itemize}

\vspace{2mm}

\subsubsection*{Vectors}

\begin{itemize}[\IEEEiedlabeljustifyl \IEEEsetlabelwidth{Z} \labelsep 0.9cm]
\item[$\bfg 0_n$] null vector of order $n$ %
\item[$\bfg A$] vector potential for solenoidal field %
\item[$\unitvec{i}$] unit vector of coordinate basis
\item[$\bfg x$] position  
\item[$\bfg v$] voltage (generalized velocity)
\item[$\vort{}$] vorticity
\item[$\bfg \varphi$] magnetic flux (generalized position)
\item[$\bfg \omega_t$] rotation due to local time dependency %
\item[$\bfg \omega_r$] rotation due to shear strain distortion %
\item[$\bfg \omega_v$] azimuthal frequency (curvature) of the voltage
\end{itemize}

\vspace{2mm}

\subsubsection*{Bivectors}

\begin{itemize}[\IEEEiedlabeljustifyl \IEEEsetlabelwidth{Z} \labelsep 0.9cm]
\item[$\hat{\bfg 0}_n$] null bivector of order $n$ %
\item[$\Wbiv{a}$] geometric frequency bivector %
\end{itemize}

\vspace{2mm}

\subsubsection*{Multivectors}

\begin{itemize}[\IEEEiedlabeljustifyl \IEEEsetlabelwidth{Z} \labelsep 0.9cm]
\item[$\hat{A}$] generic multivector, $\hat{A} = a + \hat{\bfb A}$ %
\item[$\Wgeom{a}$] geometric frequency multivector %
\end{itemize}

\vspace{2mm}

\subsubsection*{Tensors and Matrices}

\begin{itemize}[\IEEEiedlabeljustifyl \IEEEsetlabelwidth{Z} \labelsep 0.9cm]
\item[$\bfb 0_{n,m}$] null matrix of dimensions $(n, m)$ %
\item[$\bfb C$] Clarke transform matrix %
\item[$\bfb D$] pure strain tensor %
\item[$\bfb I_{n}$] identity matrix of order $n$ %
\item[$\bfb J$] Jacobian matrix %
\item[$\bfb R$] shear strain tensor %
\item[$\bfb S$] normal strain tensor %
\item[$\bfb Q$] rigid-body rotation tensor %
\end{itemize}

\vspace{2mm}

\subsubsection*{Operators}

\begin{itemize}[\IEEEiedlabeljustifyl \IEEEsetlabelwidth{Z} \labelsep 0.9cm]
\item[$T$] matrix transpose %
\item[$\dt$] partial time derivative, $\dt = \frac{\partial}{\partial t}$ %
\item[$\nabla$] nabla operator, $\nabla = (\frac{\partial}{\partial x_1}, \frac{\partial}{\partial x_2}, \dots, \frac{\partial}{\partial x_n})$ %
\item[$'$] total (Lagrange) time derivative $' = \frac{d}{dt}$ %
\item[$*$] conjugate, $\hat{A}^* = a - \hat{\bfb A}$  %
\item[$\cdot$] inner product %
\item[$\times$] cross product %
\item[$\wedge$] outer (wedge) product %
\item[$\star$] Hodge operator, $\star \hat{\bfb A} = \bfg a$ %
\end{itemize}

\section{Geometric Frequency}
\label{sec:frequency}

This section recalls the concept of geometric frequency firstly introduced for voltages and currents in \cite{freqgeom} and then generalized to any quantity in \cite{instapower}.

Consider a smooth (i.e., differentiable) time-dependent vector in a $n$-dimensional space:
\begin{equation}
  \bfg a(t) = (a_1(t), a_2(t), \dots, a_n(t)) \, ,
\end{equation}
with time derivative:
\begin{equation}
  \frac{d}{dt} \bfg a = \bfg a' = (a'_1(t), a'_2(t), \dots, a'_n(t)) \, .
\end{equation}
The \textit{geometric frequency} of $\bfg a$ is defined as the following multivector:
\begin{equation}
  \label{eq:Wgeom}
  \Wgeom{a} = \varrho_a + \Wbiv{a} =
  \frac{\bfg a \cdot \bfg a'}{|\bfg a|^2} +
  \frac{\bfg a \wedge \bfg a'}{|\bfg a|^2} \, ,
\end{equation}
which has the dimension of a frequency and is composed of a translation (scalar $\varrho_a$) and a rotation term (bivector $\Wbiv{a}$).  
$\Wgeom{a}$ is the sum of the directional and rotational derivatives of a vector along its own direction.  

If multiplied by the vector $\bfg a$ itself, $\Wgeom{a}$ returns the total time derivative of $\bfg a$, as follows:
\begin{equation}
  \label{eq:derivative0}
  \begin{aligned}
    \bfg a' &= \Wgeom{a}^* \cdot \bfg a \\
            &= \varrho_a \, \bfg a - \Wbiv{a} \cdot \bfg a \, ,
  \end{aligned}
\end{equation}
where $*$ indicates the conjugate operation, namely $\Wgeom{a}^* = \varrho_a - \Wbiv{a}$.  The term $\varrho_a \, \bfg a$ is parallel to $\bfg a$ as $\varrho_a$ is a scalar quantity.  On the other hand, the term $\Wbiv{a} \cdot \bfg a$ is perpendicular to $\bfg a$, in fact:
\begin{equation}
  \begin{aligned}
    - \Wbiv{a} \cdot \bfg a = \bfg a \cdot \Wbiv{a}
    &= \frac{1}{|\bfg a|^2} [\bfg a \cdot (\bfg a \wedge \bfg a')] \\
    &= \frac{1}{|\bfg a|^2} [(\bfg a \cdot \bfg a) \, \bfg a' - (\bfg a \cdot \bfg a') \, \bfg a] \, ,
  \end{aligned}
\end{equation}
and, hence:
\begin{equation}
  (\bfg a \cdot \Wbiv{a}) \cdot \bfg a = \frac{1}{|\bfg a|^2} [(\bfg a \cdot \bfg a)(\bfg a' \cdot \bfg a) - (\bfg a \cdot \bfg a')(\bfg a \cdot \bfg a)] = 0 \, ,
\end{equation}
where we have utilized the property that the inner product of vectors is commutative.
In three dimensions, one can write:
\begin{equation}
  \bfg \omega_a = \star \Wbiv{a} \, ,
\end{equation}
where $\star$ denotes the Hodge map \cite{Jancewicz:1989}, and
\begin{equation}
  \bfg a \times \bfg \omega_a = - \bfg a \cdot \Wbiv{a} = \Wbiv{a} \cdot \bfg a  \, .
\end{equation}
Hence, in three dimensions, \eqref{eq:derivative0} can be rewritten as:
\begin{equation}
  \label{eq:frenet0}
  \bfg a' = \varrho_a \, \bfg a + \bfg \omega_a \times \bfg a \, . 
\end{equation}

In the reminder of this work, it is assumed that a voltage at any given point of a multi-phase circuit is expressed as:
\begin{equation}
  \label{eq:voltage}
  \bfg v(t) = (v_1(t), v_2(t), \dots, v_n(t)) \, .
\end{equation}
Hence, assuming $\bfg a = \bfg v$ represents a speed of a point along a space curve, one can rewrite \eqref{eq:derivative0} as:
\begin{equation}
  \label{eq:derivative}
  \bfg v' = \Wgeom{v}^* \cdot \bfg v 
  = \varrho_v \, \bfg v - \Wbiv{v} \cdot \bfg v \, ,
\end{equation}
which, in three dimensions, becomes:
\begin{equation}
  \label{eq:frenet}
  \bfg v' = \varrho_v \, \bfg v + \bfg \omega_v \times \bfg v \, . 
\end{equation}
where $\bfg \omega_v$ is a vector the magnitude of which is proportional to the curvature of the space curve with velocity $\bfg v$ at that point and direction equal to the binormal of the Frenet frame at that point.  Moreover, $\varrho_v$ and $|\bfg \omega_v|$ are invariants and correspond, respectively, to the \textit{radial frequency}, represented by a normal strain, and \textit{azimuthal frequency}, represented by a local rotation, of a curve \cite{freqfrenet}.

\section{Lagrange Derivative}
\label{sec:lagrange}

The Lagrange (or \textit{convective} or \textit{material}) derivative expresses the time derivative of a quantity as a function of the velocity and the gradient of the vector itself.  Let us consider again a generic vector $\bfg a$ and assume that $\bfg a$ is a function of time and of time-dependent coordinates $\bfg x(t)$:
\begin{equation}
  \begin{aligned}
    \bfg a(t, \bfg x(t)) 
    = (a_1(t, \bfg x(t)), a_2(t, \bfg x(t)), \dots, a_n(t, \bfg x(t))) \, ,
  \end{aligned}
\end{equation}
where
\begin{equation}
  \bfg x(t) = (x_1(t), x_2(t), \dots, x_n(t)) \, ,
\end{equation}
Then the Lagrange derivative is written as \cite{Kambe:2007}:
\begin{equation}
  \label{eq:Lagrangian}
  \bfg a' = \dt \bfg a + (\bfg v \cdot \nabla) \bfg a \, ,
\end{equation}
where $\bfg v = \bfg x'$ is the velocity vector.

In the reminder of this work, we are interested exclusively in the time derivative of the velocity itself.
Moreover, based on \cite{freqgeom}, we assume that the voltage is a generalized velocity of a curve defined by magnetic fluxes as generalized coordinates:
\begin{equation}
  \bfg v(t) = \bfg \varphi'(t) = (\varphi'_1(t), \varphi'_2(t), \dots, \varphi'_n(t)) \, , 
\end{equation}
and, in order to link to classical fluid mechanics and to the Lagrange derivative, we assume that the voltage can be expressed as a function of time and of the magnetic fluxes:
\begin{equation}
  \bfg v(t, \bfg \varphi(t)) = (v_1(t, \bfg \varphi(t)), v_2(t, \bfg \varphi(t)), \dots, v_n(t, \bfg \varphi(t))) \, .
\end{equation}
The same assumptions applies to electric currents (generalized velocities) and electric charges (generalized positions).  In the reminder of this work, for simplicity but without lack of generality, we consider only voltages.

With the assumption that the voltage is a generalized velocity, \eqref{eq:Lagrangian} is written as:
\begin{equation}
  \label{eq:Lagv}
  \bfg v' = \dt \bfg v + (\bfg v \cdot \nabla) \bfg v \, ,
\end{equation}
or, equivalently:
\begin{equation}
  \label{eq:Lagv1}
  \bfg v' = \dt \bfg v + \frac{1}{2} \nabla |\bfg v|^2 -
  (\nabla \wedge \bfg v) \cdot \bfg v \, ,
\end{equation}
where we have used the identity (see, e.g., \cite{Kambe:2007}):
\begin{equation}
  (\bfg v \cdot \nabla) \bfg v = \frac{1}{2} \nabla |\bfg v|^2 -
  (\nabla \wedge \bfg v) \cdot \bfg v \, .
\end{equation}

For $n = 3$, the expression above can be rewritten as:
\begin{equation}
  \label{eq:Lagv2}
  \bfg v' = \dt \bfg v + \frac{1}{2} \nabla |\bfg v|^2 +
  (\nabla \times \bfg v) \times \bfg v \, ,
\end{equation}
or, equivalently:
\begin{equation}
  \label{eq:Lagv3}
  \bfg v' = \dt \bfg v + \frac{1}{2} \nabla |\bfg v|^2 +
  \vort{} \times \bfg v \, ,
\end{equation}
where $\vort{} = \nabla \times \bfg v$ is called \textit{vorticity}.

To show the correspondence of the Lagrange derivative with the geometric frequency, we need first to rewrite \eqref{eq:Lagv1} in a convenient form.  With this aim, observe that:
\begin{equation}
  \label{eq:equiv1}
  \frac{1}{2} \nabla |\bfg v|^2 =
  (\nabla \bfg v) \, \bfg v =
  \bfb J^T \bfg v \, ,
\end{equation}
where $\bfb J$ is the Jacobian of $\bfg v$.
Moreover, one has:
\begin{equation}
  \label{eq:equiv2}
  \star (\nabla \wedge \bfg v) \cdot \bfg v = -(\bfb J - \bfb J^T) \, \bfg v \, ,
\end{equation}
in fact:
\begin{equation}
  \star (\nabla \wedge \bfg v) =
  \nabla \bfg v - (\nabla \bfg v)^T =
  \bfb J^T - \bfb J \, . 
\end{equation}
Merging \eqref{eq:equiv1} and \eqref{eq:equiv2} into \eqref{eq:Lagv1} gives:
\begin{equation}
  \label{eq:vprime}
  \bfg v' =
  \dt \bfg v + \bfb J^T \bfg v + (\bfb J - \bfb J^T) \, \bfg v \, .
\end{equation}
Next, we observe that, from the Toeplitz transformation, every square matrix can be uniquely decomposed into two matrices, one symmetric and one skew-symmetric, as follows:
\begin{equation}
  \label{eq:toeplitz}
  \bfb J =
  \frac{1}{2} (\bfb J + \bfb J^T) + \frac{1}{2} (\bfb J - \bfb J^T) =
  \bfb D + \bfb Q \, ,
\end{equation}
where $\bfb D$ is symmetric, i.e., $\bfb D^T = \bfb D$, and $\bfb Q$ is skew-symmetric, i.e., $\bfb Q^T = - \bfb Q$.  The symmetric part represents a \textit{pure strain motion} and $\bfb D$ is called \textit{pure strain tensor}, whereas the skew-symmetric part represents a \textit{rigid-body rotation} and $\bfb Q$ is a tensor that defines a local rigid-body angular rotation.

The symmetric term $\bfb D$ can be further split into two symmetric matrices:
\begin{equation}
  \bfb D = \bfb S + \bfb R \, ,
\end{equation}
\begin{equation}
  \label{eq:S}
  \bfb S = \frac{1}{n}\Trace{J} \, \bfb I_n \, ,
\end{equation}
where $\bfb I_n$ is the identity matrix of order $n$ and $\Trace{J}$
is the trace of matrix $\bfb J$; and:
\begin{equation}
  \label{eq:R}
  \bfb R = \bfb D - \bfb S \, .
\end{equation}
In fluid mechanics, $\bfb S$ and $\bfb R$ are called \textit{normal strain tensor} and \textit{shear strain tensor}, respectively.  The normal strain tensor represents either an expansion or a compression and is perpendicular to the curve $\bfg x(t)$.  The shear strain tensor represents a distortion and is parallel to the curve $\bfg x(t)$.

From \eqref{eq:equiv2}, and the definition of tensor $\bfb Q$ in \eqref{eq:toeplitz}, one obtains:
\begin{equation}
  \label{eq:Qstar}
  \bfb Q = \frac{1}{2} \nabla \wedge \bfg v \, ,
\end{equation}
and, in three dimensions:
\begin{equation}
  \star \bfb Q = \frac{1}{2} \nabla \times \bfg v = \frac{1}{2} \, \vort{} \, .
\end{equation}

Tensor $\bfb S$ is diagonal and have all diagonal elements equal to each other.  It represents velocities that are purely radial, in fact:
\begin{equation}
  \label{eq:trace}
  \Trace{J} = \sum_{i=1}^n J_{ii} =
  \sum_{i=1}^n \frac{\partial v_i}{\partial x_i} =
  \Div{v} \, ,
\end{equation}
that is, the diagonal elements are proportional to the divergence of the velocity vector.

Using the decomposition above, $\bfb J$ and $\bfb J^T$ are written as:
\begin{equation}
  \label{eq:SRQ}
  \bfb J = \bfb S + \bfb R + \bfb Q \quad \Rightarrow \quad
  \bfb J^T = \bfb S + \bfb R - \bfb Q \, ,
\end{equation}
and substituting the expression of $\bfb J^T$ in \eqref{eq:vprime}, one has:
\begin{equation}
  \label{eq:SRQv}
  \bfg v' = \dt \bfg v + \bfb S \, \bfg v + \bfb R \, \bfg v + \bfb Q \, \bfg v \, .
\end{equation}

\section{Equivalence between \eqref{eq:derivative} and \eqref{eq:Lagv1}}
\label{sec:equivalence}

We are now ready to show the equivalence between \eqref{eq:derivative} and \eqref{eq:Lagv1}.  Substituting the expression of $\bfg v'$ from \eqref{eq:SRQv} in the expression of the geometric frequency multivector \eqref{eq:Wgeom} leads to:
\begin{equation}
  \label{eq:Wv2full}
  \begin{aligned}
    \Wgeom{v}
    &=
    \frac{\bfg v \cdot \dt \bfg v}{|\bfg v|^2} +
    \frac{\bfg v \wedge \dt \bfg v}{|\bfg v|^2} +
    \frac{\bfg v^T (\bfb S \bfg v)}{|\bfg v|^2} +
    \frac{\bfg v \wedge (\bfb S \bfg v)}{|\bfg v|^2} \\
    &+
    \frac{\bfg v^T (\bfb R \bfg v)}{|\bfg v|^2} +
    \frac{\bfg v \wedge (\bfb R \bfg v)}{|\bfg v|^2} +
    \frac{\bfg v^T (\bfb Q \bfg v)}{|\bfg v|^2} + 
    \frac{\bfg v \wedge (\bfb Q \bfg v)}{|\bfg v|^2} \, .
  \end{aligned}
\end{equation}
This expression can be simplified as discussed next.  First, observe that, by construction of $\bfb Q$, one has:
\begin{equation}
  \label{eq:id1}
  \bfg v^T (\bfb Q \bfg v) = \frac{1}{2} (\bfg v^T (\bfb J \bfg v) - \bfg v^T (\bfb J^T \bfg v)) = 0 \, ,
\end{equation}
in fact
\begin{equation}
  \bfg v^T (\bfb J \bfg v) = (\bfg v^T (\bfb J \bfg v))^T = \bfg v^T (\bfb J^T \bfg v) \, .
\end{equation}

Then observe that $\bfb S$ is diagonal and all its diagonal elements are equal.  Using \eqref{eq:S} and \eqref{eq:trace}, one has:
\begin{equation}
  \bfg v \wedge (\bfb S \bfg v) = \frac{1}{n} \Trace{J} (\bfg v \wedge (\bfb I_n \bfg v)) = \frac{1}{n} (\Div{v}) (\bfg v \wedge \bfg v) = \bivec{0}_n \, ,
\end{equation}
where we have also utilized the identity $\bfg v = \bfb I_n \, \bfg v$ and the property of the outer product $\bfg  a \wedge \bfg a = \bivec{0}_n$.  Hence:
\begin{align}
  \label{eq:id2}
  \bfg v \wedge (\bfb S \bfg v) = \bivec{0}_n \, ,
\end{align}
where $\bivec{0}_n$ is the null bivector of order $n$.  Then, \eqref{eq:Wv2full} can be rewritten as:
\begin{equation}
  \label{eq:Wv2}
  \begin{aligned}
    \Wgeom{v}
    &=
    \frac{\bfg v \cdot \dt \bfg v}{|\bfg v|^2} +
    \frac{\bfg v^T (\bfb S \bfg v)}{|\bfg v|^2} +
    \frac{\bfg v^T (\bfb R \bfg v)}{|\bfg v|^2} \\
    &+
    \frac{\bfg v \wedge \dt \bfg v}{|\bfg v|^2} +
    \frac{\bfg v \wedge (\bfb R \bfg v)}{|\bfg v|^2} +
    \frac{\bfg v \wedge (\bfb Q \bfg v)}{|\bfg v|^2} \, .
  \end{aligned}
\end{equation}

Under the conditions of the Helmoltz-Zorawski criterion, also the following identity holds:
\begin{equation}
  \label{eq:id0}
  \bfg v^T \wedge \dt \bfg v = \bivec{0}_n \, ,
\end{equation}
The Helmoltz-Zorawski criterion states that a necessary and sufficient condition for the lines of a vectorial field $\bfg a(t, \bfg x(t))$ to be a material curves is that \cite{Petrila:2005}:
\begin{equation}
  \label{eq:helmoltz}
  \bfg a \times \left [ \dt \bfg a + \nabla \times (\bfg a \times \bfg v) + \bfg v\nabla \cdot \bfg a \right ] = \bfg 0 \, ,
\end{equation}
then if $\bfg a = \bfg v$, \eqref{eq:helmoltz} reduces to the identity \eqref{eq:id0}.
In \cite{Stern:1966}, it is shown that \eqref{eq:helmoltz} -- and hence \eqref{eq:id0} -- is satisfied if the vectorial field satisfies the condition of \textit{flux conservation} \cite{Truesdell:1954}.  In the examples, we show that flux conservation is satisfied for most operating conditions of voltages (currents).  A relevant exception is the case of stationary voltages with harmonics.

Using again \eqref{eq:S} and \eqref{eq:trace}, one can observe that:
\begin{equation}
  \frac{\bfg v^T (\bfb S \bfg v)}{|\bfg v|^2} =
  \frac{1}{n} \Trace{J} \frac{\bfg v^T \bfb I_n \bfg v}{|\bfg v|^2} =
  \frac{1}{n} \Trace{J} = \frac{1}{n} \Div{v} \, .
\end{equation}

Equaling the scalar and the multivector terms of \eqref{eq:derivative} and \eqref{eq:Wv2}, one obtains:
\begin{align}
  \label{eq:rhoL}
  \varrho_v &= \frac{\bfg v \cdot \dt \bfg v}{|\bfg v|^2} +
  \frac{1}{n} \Div{v} + \frac{\bfg v^T (\bfb R \bfg v)}{|\bfg v|^2} \, , \\
  \label{eq:WL}
  \Wbiv{v} &= \frac{\bfg v \wedge \dt \bfg v}{|\bfg v|^2} +
  \frac{\bfg v \wedge (\bfb R \bfg v)}{|\bfg v|^2} +
  \frac{\bfg v \wedge (\bfb Q \bfg v)}{|\bfg v|^2} \, .
\end{align}
and substituting \eqref{eq:Qstar} into \eqref{eq:WL}:
\begin{equation}
  \label{eq:Wv3}
  \Wbiv{v} = \frac{\bfg v \wedge \dt \bfg v}{|\bfg v|^2} +
  \frac{\bfg v \wedge (\bfb R \bfg v)}{|\bfg v|^2} +
  \frac{1}{2} \frac{\bfg v \wedge (\star (\nabla \wedge \bfg v) \cdot \bfg v)}{|\bfg v|^2} \, .
\end{equation}
In three dimensions, \eqref{eq:Wv3} can be written as:
\begin{equation}
  \label{eq:Wv4}
  \bfg \omega_v = \frac{\bfg v \times \dt \bfg v}{|\bfg v|^2} +
  \frac{\bfg v \times (\bfb R \bfg v)}{|\bfg v|^2} +
  \frac{1}{2} \frac{\bfg v \times (\vort{} \times \bfg v)}{|\bfg v|^2} \, ,
\end{equation}
From the Lagrange's formula of triple vector product, one has:
\begin{equation}
  \bfg v \times (\vort{} \times \bfg v) =
  (\bfg v \cdot \bfg v) \vort{} - (\vort{} \cdot \bfg v) \bfg v =
  |\bfg v|^2 \vort{} \, ,
\end{equation}
where $\vort{} \cdot \bfg v = 0$ as, for the definition of the the curl of a vector, the vorticity is perpendicular to the velocity.  Equation \eqref{eq:Wv4} can be thus written as:
\begin{equation}
  \label{eq:Wvw}
  \bfg \omega_v = \frac{\bfg v \times \dt \bfg v}{|\bfg v|^2} +
  \frac{\bfg v \times (\bfb R \bfg v)}{|\bfg v|^2} + \frac{1}{2} \vort{} \, .
\end{equation}

The special case, in three dimensions, that satisfies $\dt \bfg v = \bfg 0$, which implies stationarity, and presents no distortion, i.e., $\bfb R = \bfg 0_{3, 3}$, gives:
\begin{equation}
  \label{eq:noR}
  \varrho_v = \frac{1}{3} \Div{v} \, , \qquad
  \bfg \omega_v = \frac{1}{2} \vort{} = \frac{1}{2} \nabla \times \bfg v \, .
\end{equation}

\section{Examples}
\label{sec:examples}

In this section we consider typical stationary ac cases, namely balanced and unbalanced, sinusoidal and non-sinusoidal, as well a time-varying dc case.  All ac cases are based on the Clarke's transform of a three-phase voltage vector (see the Appendix).  This simplifies the notation and makes simpler visualizing the results while, at the same time, no information is lost.

To ease the notation of the examples discussed below, we introduce the following quantities:
\begin{equation*}
  \begin{aligned}
    \rho_t &= \frac{\bfg v \cdot \dt \bfg v}{|\bfg v|^2} \, ,
    &\qquad
     \bfg \omega_t &=  \frac{\bfg v \times \dt \bfg v}{|\bfg v|^2} \, , \\
    \rho_r &= \frac{\bfg v^T (\bfb R \bfg v)}{|\bfg v|^2} \, ,
    &\qquad
    \bfg \omega_r &= \frac{\bfg v \times (\bfb R \bfg v)}{|\bfg v|^2} \, .
  \end{aligned}
\end{equation*}

\subsection{Stationary balanced sinusoidal ac voltage}

We consider first a stationary balanced sinusoidal voltage of a three-phase circuit.  In this scenario, the Clarke components of the voltage are given by:
\begin{equation*}
  \begin{aligned}
    \bfg v(t) &= (v_{\alpha}(t) \, , v_{\beta}(t) \, , v_\gamma(t)) \\
    &= ( V \cos (\omega_o t + \phi) \, , V \sin (\omega_o t + \phi), 0) \\
    &= v_{\alpha}(t) \, \unitvec{\alpha} + v_{\beta}(t) \, \unitvec{\beta} \, ,
  \end{aligned}
\end{equation*}
and the $v_{\gamma} = 0$ as the voltages are balanced.
The geometric frequency of $\bfg v(t)$ gives (see also a similar example in \cite{freqfrenet}):
\begin{equation*}
  \begin{aligned}
    \rho_v &= \frac{v_{\alpha} v'_{\alpha} + v_{\beta} v'_{\beta}}{V^2} = 0 \, , \\
    \bfg \omega_v &= \frac{v_{\alpha} v'_{\beta} - v_{\beta} v'_{\alpha}}{V^2} \,
    (\unitvec{\alpha} \times \unitvec{\beta})
    = \omega_o \, \unitvec{\gamma} \, ,
  \end{aligned}
\end{equation*}
where we have used the identities $\star (\unitvec{\alpha} \wedge \unitvec{\beta}) = \unitvec{\alpha} \times \unitvec{\beta} = \unitvec{\gamma}$.  $\unitvec{\gamma}$ is the unit binormal in the Clarke transform space $(\alpha, \beta, \gamma)$ and is perpendicular to the $(\alpha, \beta)$ plane.  As $\rho_v$ and $|\bfg \omega_v|$ are geometrical invariants, the same results can be obtained using any other coordinate system, hence including the original $abc$ voltages that can be measured at the phases of the physical circuit.

Using the notation common to fluid mechanics, one can rewrite the voltage vector as a function of the fluxes, namely, $\bfg v(\bfg \varphi(t))$, where:
\begin{equation*}
  \begin{aligned}
    \bfg \varphi(t)
    &= (\varphi_{\alpha}(t) \, , \varphi_{\beta}(t) \, ,  \varphi_{\gamma}(t) ) \\
    &= \left ( \frac{V}{\omega_o} \sin (\omega_o t + \phi) \, ,
    -\frac{V}{\omega_o} \cos (\omega_o t + \phi) \, , 0 \right ) \\
    &= \varphi_{\alpha}(t) \, \unitvec{\alpha} + \varphi_{\beta}(t) \, \unitvec{\beta} \, ,
  \end{aligned}
\end{equation*}
where we have assumed that the fluxes have no constant terms, i.e., the circuit does not include fixed permanent magnets.  Then the voltage can be written as:
\begin{equation*}
  \bfg v(\bfg \varphi(t)) = -\omega_o \varphi_{\beta}(t) \, \unitvec{\alpha} +
  \omega_o \varphi_{\alpha}(t) \, \unitvec{\beta} \, ,
\end{equation*}
and the divergence and curl of the voltage are:
\begin{equation*}
  \begin{aligned}
    \nabla \cdot \bfg v &=
    \frac{\partial v_{\alpha}}{\partial \varphi_{\alpha}} +
    \frac{\partial v_{\beta}}{\partial \varphi_{\beta}} = 0 \, , \\
    \nabla \times \bfg v &=
    \left ( \frac{\partial v_{\beta}}{\partial \varphi_{\alpha}} -
    \frac{\partial v_{\alpha}}{\partial \varphi_{\beta}} \right ) \unitvec{\gamma}
    = 2\omega_o \, \unitvec{\gamma} \, , 
  \end{aligned}
\end{equation*}
which indicates that the vorticity has magnitude $2\omega_o$.  Then, as the voltage is stationary, $\dt \bfg v = \bfg 0_3$.  It remains to determine the distortion of the strain and of rotation due to the shear strain tensor $\bfb R$.  First, we note that, as $\nabla \cdot \bfg v=0$, $\bfb S = \bfb 0_{3,3}$ and, hence, $\bfb R = \bfb D$.  Then, we observe that also $\bfb D = \bfb 0_{3,3}$, in fact, $\bfb J$ is skew-symmetric:
\begin{equation*}
  \bfb J = 
  \begin{bmatrix}
    0 & \frac{\partial v_{\alpha}}{\partial \varphi_{\beta}} & 0 \\
    \frac{\partial v_{\beta}}{\partial \varphi_{\alpha}} & 0 & 0 \\
    0 & 0 & 0 \\
  \end{bmatrix} =
  \begin{bmatrix}
    0 & -\omega_o & 0 \\
    \omega_o & 0 & 0 \\
    0 & 0 & 0 \\
  \end{bmatrix} = \bfb Q \, .
\end{equation*}
Substituting the results above in \eqref{eq:rhoL} and \eqref{eq:Wv4}, one obtains that $\rho_v = 0$ and $\bfg \omega_v = \frac{1}{2} \vort{} = \omega_o \, \unitvec{\gamma}$, as expected.

In field theory, this condition makes a field \textit{solenoidal}.  We can thus say that, using this analogy, $\bfg v$ is a solenoidal vector field that is obtained from a vector potential, say $\bfg A$, that satisfies the condition $\bfg v = \nabla \times \bfg A$.  Moreover, in fluid mechanics, a fluid for which $\nabla \cdot \bfg v = 0$ applies everywhere is \textit{incompressible} as this condition implies that the volume of a particle of the fluid cannot vary \cite{Kambe:2007}.  This suggests that an ac electric circuit in stationary conditions, can be studied as an incompressible fluid.

\subsection{DC voltage}

We consider a dc voltage and proceed similarly to what done in \cite{freqgeom}.  In dc circuits, the voltage has only one component along the single coordinate of the system, say $\unitvec{\rm dc}$, hence:
\begin{equation*}
  \bfg v(t) = v_{\rm dc}(t) \, \unitvec{\rm dc} \, , \qquad
  \bfg v'(t) = v_{\rm dc}'(t) \, \unitvec{\rm dc} \, ,
\end{equation*}
and, from the definition of geometric frequency:
\begin{equation*}
  \begin{aligned}
    \rho_v(t) &= \frac{\bfg v(t) \cdot \bfg v'(t)}{v^2_{\rm dc}(t)} = \frac{v_{\rm dc}'(t)}{v_{\rm dc}(t)} \, , \\
    \bfg \omega_v &= \frac{\star (\bfg v(t) \wedge \bfg v'(t))}{v^2_{\rm dc}(t)} =
    \frac{v_{\rm dc}'(t)}{v_{\rm dc}(t)} \,
    (\unitvec{\rm dc} \times \unitvec{\rm dc}) = \bfg 0_1 \, .
  \end{aligned}
\end{equation*}
Following the approach based on the Lagrange derivative, rewriting the voltage vector as $\bfg v = v_{\rm dc}(t, \varphi_{\rm dc}(t)) \, \unitvec{\rm dc}$, and substituting the terms in \eqref{eq:rhoL}, one has:
\begin{equation}
  \label{eq:rhodc}
  \begin{aligned}
    \rho_v(t) = \frac{\dt v_{\rm dc}(t)}{v_{\rm dc}(t)} +
    \frac{\partial v_{\rm dc}(t)}{\partial \varphi_{\rm dc}(t)} \, .
  \end{aligned}
\end{equation}
where we used the fact that $\bfb R = \bfb 0_{1,1}$ as $\bfb J = \bfb D = \frac{\partial v_{\rm dc}(t)}{\partial \varphi_{\rm dc}(t)}$.  Then, it descends that also $\bfb Q = \bfb 0_{1,1}$ and, hence, $\bfg \omega_v = \bfg 0_1$.  This last condition implies that $\nabla \times \bfg v = \bfg 0_1$ and hence, the voltage is an \textit{irrotational} field, which can be obtained from a scalar potential, say $\tilde{\varphi}(t)$, which satisfies the condition $\bfg v(t) = \nabla \tilde{\varphi}(t)$.  But, in one dimension, one has:
\begin{equation*}
  \bfg v(t) =
  \nabla \tilde{\varphi}(t) =
  \nabla \cdot (\tilde{\varphi}(t) \, \unitvec{\rm dc}) \, ,
\end{equation*}
which leads to rewrite \eqref{eq:rhodc} as:
\begin{equation*}
  \begin{aligned}
    \rho_v(t) = 
    \frac{d v_{\rm dc}(t)}{d \varphi_{\rm dc}(t)} \, .
  \end{aligned}
\end{equation*}
where $\varphi_{\rm dc}(t) \equiv \tilde{\varphi}(t)$, that is, the potential of the voltage is the magnetic flux, as expected.  This result is also consistent with Faraday's law and the chain rule:
\begin{equation}
  \frac{d v_{\rm dc}(t)}{dt} =
  \frac{d v_{\rm dc}(t)}{d \varphi_{\rm dc}(t)} \,
  \frac{d \varphi_{\rm dc}(t)}{d t} =
  \rho_v(t) \, v_{\rm dc} (t) \, .
\end{equation}

\subsection{Stationary unbalanced sinusoidal ac voltage}

We consider now a stationary unbalanced sinusoidal voltage of a three-phase circuit.  Let us assume that the Clarke components of the voltage are given by:
\begin{equation*}
  \begin{aligned}
    \bfg v(t) &= ( V_{\alpha} \cos (\omega_o t + \phi) \, ,
    V_{\beta} \sin (\omega_o t + \phi) )\, ,
  \end{aligned}
\end{equation*}
which leads to:
\begin{equation*}
  \begin{aligned}
    \rho_v(t) &= \frac{1}{2}\frac{\omega_o (V_{\beta}^2 - V_{\alpha}^2) \sin (2\theta(t))}{|\bfg v(t)|^2} \, , \\
    \bfg \omega_v(t) &= \omega_v(t) \, \unitvec{\gamma} = \frac{\omega_o V_{\alpha} V_{\beta}}{|\bfg v(t)|^2} \, \unitvec{\gamma} \, .
  \end{aligned}
\end{equation*}
where $\theta(t) = \omega_o t + \phi$.
Using the formalism of the Lagrange derivative, we can write the fluxes as:
\begin{equation*}
  \begin{aligned}
    \bfg \varphi(t)
    &= \left ( \frac{V_{\alpha}}{\omega_o} \sin \theta(t) \, , \,
      - \frac{V_{\beta}}{\omega_o} \cos \theta(t) \, , 0 \right ) ,
  \end{aligned}
\end{equation*}
and, hence:
\begin{equation*}
  \bfg v(\bfg \varphi(t)) =
  \left ( -\omega_o \frac{V_{\alpha}}{V_{\beta}} \varphi_{\beta}(t) \, , \,
  \omega_o \frac{V_{\beta}}{V_{\alpha}} \varphi_{\alpha}(t) \, , 0 \right ) .
\end{equation*}
Then the Jacobian matrix becomes:
\begin{equation*}
  \bfb J = (\nabla \bfg v)^T =
  \begin{bmatrix}
    0 & -\omega_o \frac{V_{\alpha}}{V_{\beta}} & 0 \\
    \omega_o \frac{V_{\beta}}{V_{\alpha}} & 0 & 0 \\
    0 & 0 & 0 \\
  \end{bmatrix} ,
\end{equation*}
which can be decomposed as:
\begin{equation}
  \bfb J = \bfb R + \bfb Q \, ,
\end{equation}
where $\bfb S = \bfg 0_{3,3}$, which also implies $\nabla \cdot \bfg v = 0$ as well as solenoidal field and incompressible fluid, and:
\begin{equation*}
  \bfb R =
  \begin{bmatrix}
    0 & \kappa \omega_o & 0 \\
    \kappa \omega_o & 0 & 0 \\
    0 & 0 & 0 \\
  \end{bmatrix} , \qquad
  \bfb Q =
  \begin{bmatrix}
    0 & -\xi \omega_o & 0 \\
    \xi \omega_o & 0 & 0 \\
    0 & 0 & 0 \\
  \end{bmatrix} ,  
\end{equation*}
with
\begin{equation*}
  \kappa = \frac{1}{2} \left ( \frac{V_{\beta}}{V_{\alpha}} - \frac{V_{\alpha}}{V_{\beta}} \right )\, , \qquad
  \xi =  \frac{1}{2} \left ( \frac{V_{\alpha}}{V_{\beta}} + \frac{V_{\beta}}{V_{\alpha}} \right ) \, .
\end{equation*}
As the voltage is stationary, $\dt \bfg v = \bfg 0$, then:
\begin{equation*}
  \rho_v(t) = \rho_r(t) =
  \kappa \omega_o \frac{ V_{\alpha} V_{\beta} \sin^2 \theta(t) }{|\bfg v(t)|^2} \, ,
\end{equation*}
where $\kappa V_{\alpha} V_{\beta} = \frac{1}{2}(V^2_{\beta} - V^2_{\alpha})$ and which indicates that the unbalance in the voltage magnitudes creates a shear strain that is responsible of periodically stretching and shrinking the radius of the trajectory in order to obtain an ellipse.  Note that, as expected, the expressions of $\rho_v$ obtained using the differential geometry approach and the Lagrange derivative coincide.

The rotation, on the other hand, has two terms:
\begin{equation*}
  \bfg \omega_v(t) = \bfg \omega_r(t) + \frac{1}{2} \vort{} \, ,
\end{equation*}
where the local distortion of the rotation is:
\begin{equation*}
  \begin{aligned}
    \bfg \omega_r
    &= \omega_r \, \unitvec{\gamma} =
      \kappa \omega_o \frac{V_{\alpha}^2 \cos^2 \theta -
      V_{\beta}^2 \sin^2 \theta}{|\bfg v|^2} \, \unitvec{\gamma} \\
    &= \frac{1}{2} \omega_o \left ( \frac{V_{\alpha} V_{\beta}}{|\bfg v|^2}
    - \frac{V^3_{\alpha}}{V_{\beta}} \cos^2 \theta
    - \frac{V^3_{\beta}}{V_{\alpha}} \sin^2 \theta \right ) \unitvec{\gamma} ,
  \end{aligned}
\end{equation*}
where the dependency on time has been dropped for simplicity, and the rigid-body rotation is:
\begin{equation*}
  \begin{aligned}
    \frac{1}{2} \vort{} = \frac{\bfg v(t) \times (\bfb Q \bfg v(t))}{|\bfg v(t)|^2} 
    = \xi \omega_o \, \unitvec{\gamma} \, .
  \end{aligned}
\end{equation*}
It is interesting to observe that while the magnitude of $\bfg \omega_v$ is time varying, the magnitude of the vorticity is constant.  The angular frequency $\xi \omega_o$ can be interpreted as the constant curvature of a stationary balanced voltage and the term due to the shear strain as the distortion on top of the constant rotation.  Note also that, if $V_{\alpha} = V_{\beta}$, $\kappa = 0$ and $\xi = 1$, which leads to the results obtained in Example 1.  Moreover for unbalances for which the difference between $V_{\alpha}$ and $V_{\beta}$ is below $15\%$, $\xi - 1 < 1\%$.  This means that in usual unbalanced conditions, one can assume that the angular speed of the rigid body is approximately $\omega_o$.

\begin{figure}[htb]
  \resizebox{0.925\linewidth}{!}{\includegraphics{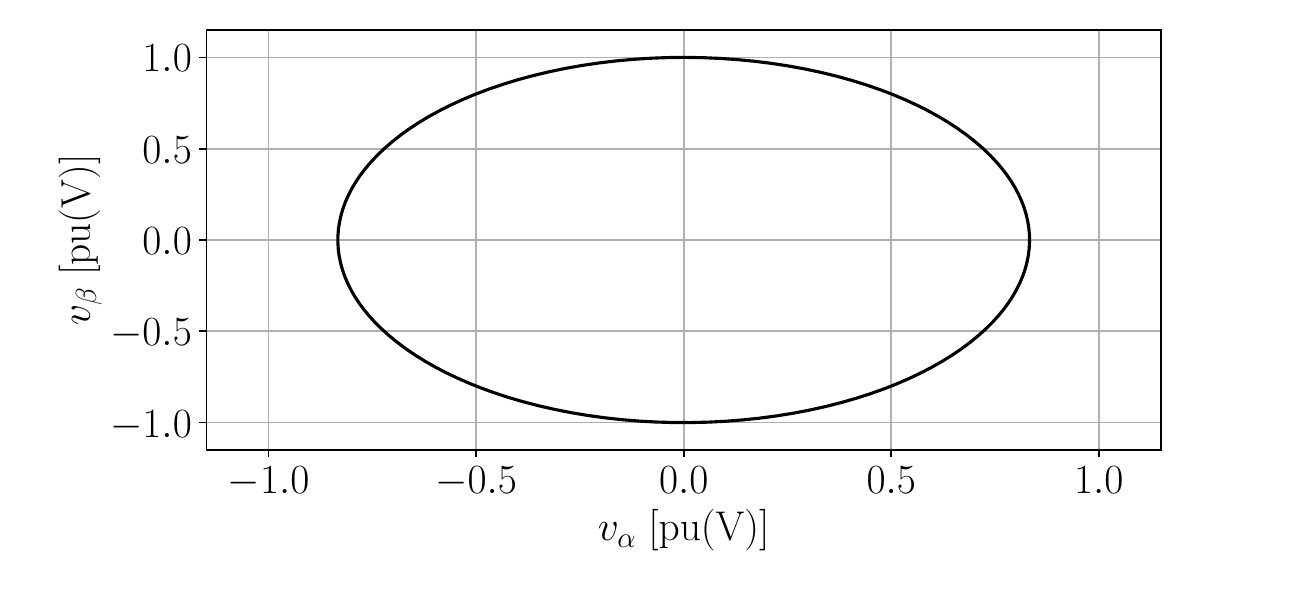}} \\
  \resizebox{0.925\linewidth}{!}{\includegraphics{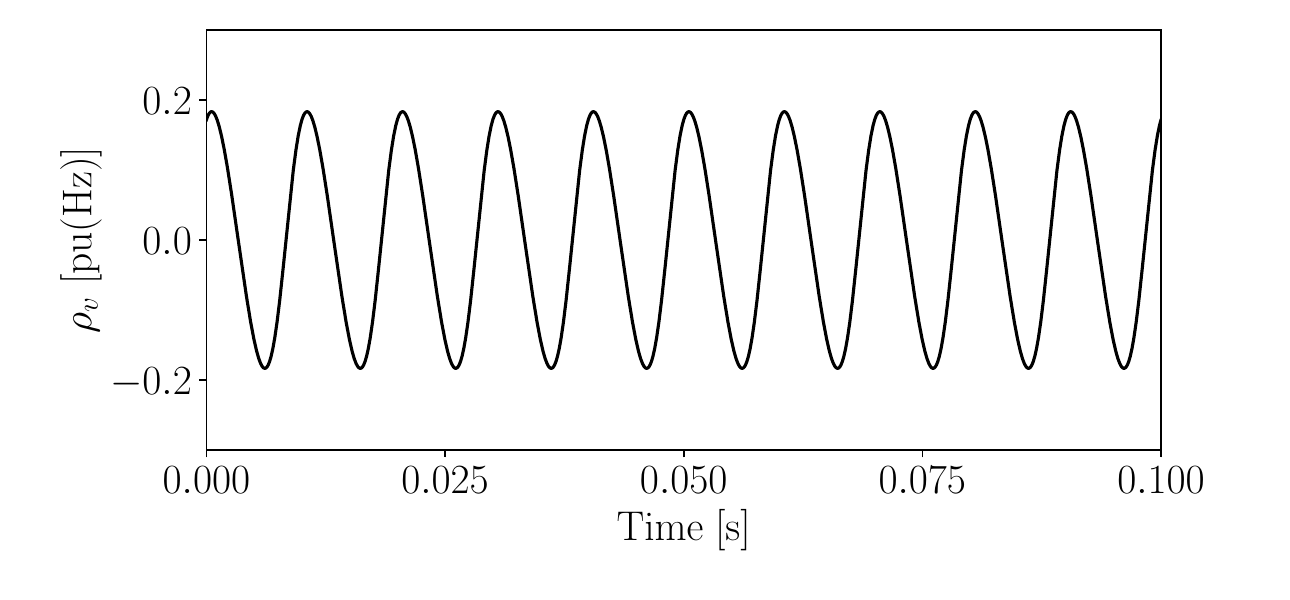}} \\
  \resizebox{0.925\linewidth}{!}{\includegraphics{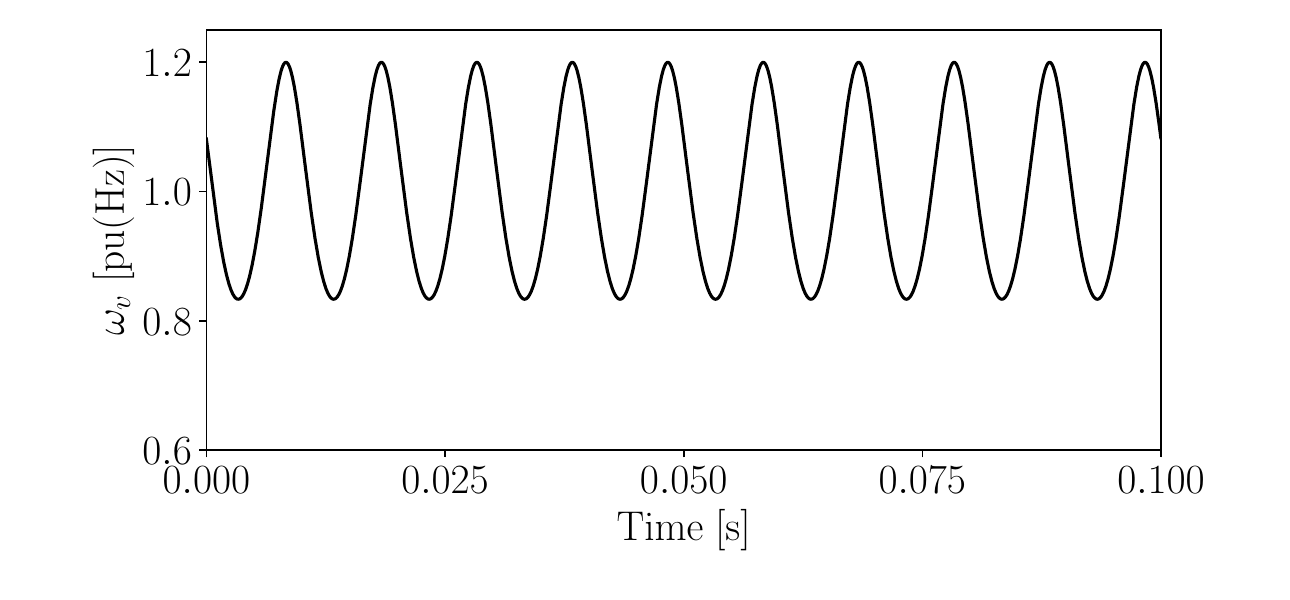}} \\
  \resizebox{0.925\linewidth}{!}{\includegraphics{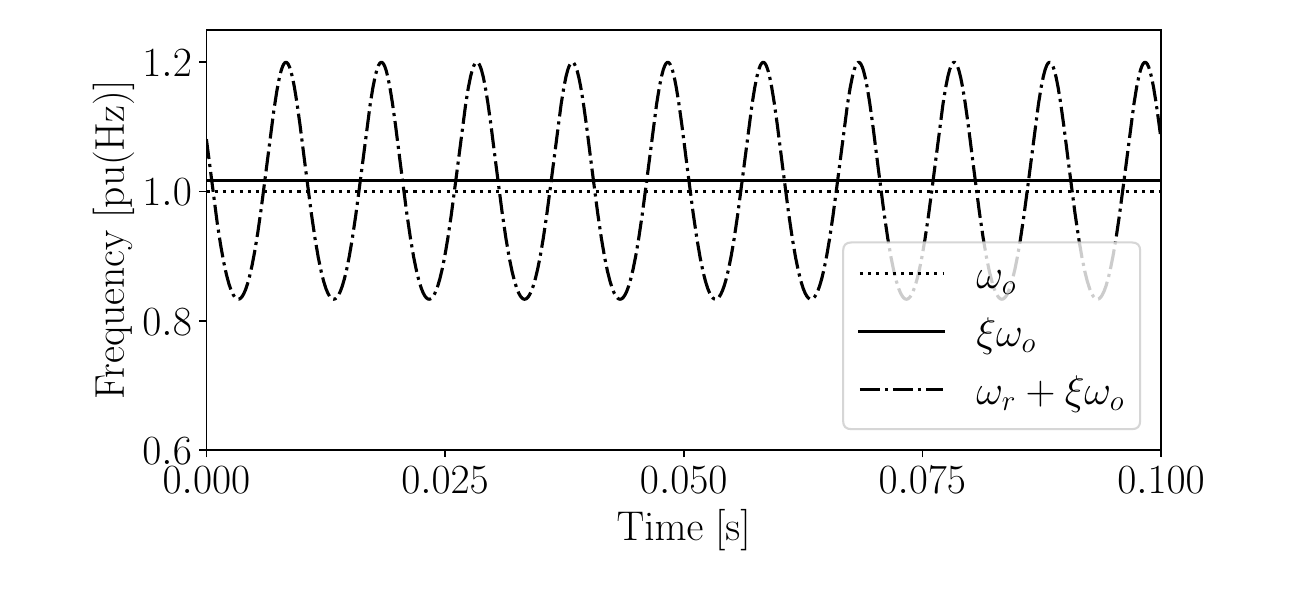}}
  \caption{Stationary unbalanced sinusoidal ac voltage and components of its geometrical frequency and Lagrange derivative.}
  \label{fig:unbalanced}
  \vspace{-3mm}
\end{figure}

Figure \ref{fig:unbalanced} illustrates an example of unbalanced voltage.  The parameters are: $V_{\beta}/V_{\alpha} = 1.2$, $\omega_o = 2\pi \, 50$ rad/s, and $\phi = \pi/6$.  Then one has:
\begin{equation*}
  \kappa = 0.1833 \, , \qquad \xi = 1.0167 \, .
\end{equation*}
Moreover, as discussed above $\rho_v(t) = \rho_r(t)$.  As expected, the figure shows that $\omega_v(t) = \omega_r(t) + \xi \omega_o$.  This means that the average of $\omega_v$ is $\xi \omega_o > \omega_o$.  This is due to the fact that the distortion $\omega_r$ is not a symmetrical function with respect to the horizontal axis.

This example demonstrates how the breakdown of the geometric frequency obtained using the Lagrange derivative and fluid mechanics formalism allows a natural separation between various components with precise geometrical meaning, as follows:
\begin{enumerate}[(i)]
\item Half the magnitude of the vorticity, namely $\xi \omega_o$, is the equivalent (virtual) frequency of the balanced voltage after removing the distortion.
\item $\omega_r$ is the distortion of the frequency due to the unbalance.
\item $\rho_r$ indicates the radial distortion (stretching/shrinking) of the radius of the voltage curve (in this case, an ellipse) in the $(\alpha, \beta)$ plane.
\end{enumerate}

The scenario discussed in this section is a special case where $v_{\gamma} = 0$, which requires that the sum of phase voltages satisfies the condition $v_a + v_b + v_c = 0$, $\forall t$.  There exist, of course, unbalanced cases for which $v_{\gamma} \ne 0$.  However, the curve associated to stationary unbalanced sinusoidal three-phase voltage is a plane curve, as shown in \cite{freqfrenet}.  Thus, the example discussed in this section has general validity, as one can always find a transformation of the $abc$ coordinates for which one component is null for all $t$ and the curve in the other two coordinates is an ellipse.

\subsection{Stationary balanced non-sinusoidal ac voltage}

Stationary balanced non-sinusoidal ac voltages allow showcasing the added value of the proposed analogy with stream lines and Lagrange derivative.  Without lack of generality, we consider the case of a voltage with two harmonics.  As done in the previous examples, we use the Clarke transform in order to reduce the coordinates to a two-dimensional space:
\begin{equation*}
  \begin{aligned}
  \bfg v(t) &=
  \begin{bmatrix}
    v_{\alpha}(t) \\
    v_{\beta}(t) \\
    v_{\gamma}(t)
  \end{bmatrix}
    =
  \begin{bmatrix}
    V \cos \theta(t) + V_h \cos \theta_h(t) \\
    V \sin \theta(t) + V_h \sin \theta_h(t) \\
    0
  \end{bmatrix} ,
  \end{aligned}
\end{equation*}
where
\begin{equation*}
  \begin{aligned}
    \theta(t) &= \omega_o t + \phi \, , \\
    \theta_h(t) &= h\omega_o t + \phi_h \, ,
  \end{aligned}
\end{equation*}
and where all parameters are constant, $h$ is typically an odd integer $> 3$, and from where one can obtain the magnetic flux vector:
\begin{equation*}
  \begin{aligned}
  \bfg \varphi(t) =
  \begin{bmatrix}
    \varphi_{\alpha}(t) \\
    \varphi_{\beta}(t) \\
    \varphi_{\gamma}(t)    
  \end{bmatrix}
  =
  \begin{bmatrix}
    \Phi \sin \theta(t) + \Phi_h \sin \theta_h(t) \\
    -\Phi \cos \theta(t) - \Phi_h \cos \theta_h(t) \\
    0
  \end{bmatrix} ,
  \end{aligned}
\end{equation*}
where $\Phi = V/\omega_o$ and $\Phi_h = V_h/(h \omega_o)$.  The
voltages can be rewritten as follows:
\begin{equation*}
  \bfg v(t, \bfg \varphi(t)) =
  \begin{bmatrix}
    -\omega_o \varphi_{\beta}(t) - \zeta_h \, V_h \cos \theta_h(t) \\
    \omega_o \varphi_{\alpha}(t) - \zeta_h \, V_h \sin \theta_h(t) \\
    0
  \end{bmatrix} ,
\end{equation*}
where $\zeta_h = (h-1)/h$.  The extension to systems with multiple harmonics is readily obtained with a summation of $h$-terms.

It is relevant to observe that, despite being considered in the literature of harmonic analysis as a ``stationary'' set of voltages, the voltages are not actually stationary from the point of view of a stream line of a fluid flow.  This is only an apparent inconsistency.  From the harmonic analysis point of view, the voltage is stationary as its Fourier transform gives crisp constant spectrum parameters.  The voltage is thus stationary in frequency domain.  From the fluid mechanics point of view, on the other hand, which is in time domain, the voltage is \textit{not} stationary as the velocity (voltage) cannot be described only in terms of its position coordinates (magnetic flux).  Thus, in this case, $\dt \bfg v \ne \bfg 0$.  Moreover, one has:
\begin{equation*}
  \begin{aligned}
    \bfg v \cdot \dt \bfg v &= -(h-1)(VV_h \sin(\theta + \theta_h) + V^2_h \sin (2\theta_h) \, ,\\
    \bfg v \times \dt \bfg v &= -(h-1)(VV_h \cos(\theta+\theta_h) + V_h^2 \cos(2\theta_h)) \, \unitvec{\gamma} \, ,
  \end{aligned}
\end{equation*}
where time dependency has been omitted for simplicity of notation.  The latter expression indicates that the curve is not a material curve as the Helmoltz-Zorawski criterion -- see \eqref{eq:id0} -- is not satisfied in this case.

Then, similarly to Examples 1 and 3, one has $\bfb S = \bfb R = \bfb 0_{3,3}$ and
$\nabla \cdot \bfg v = 0$,
which means that the voltage can be associated to an incompressible fluid and a solenoidal field.  Then, one has:
\begin{equation*}
  \rho_v(t) = \rho_t(t) \, , \qquad \bfg \omega_v(t) = \bfg \omega_t(t) + \frac{1}{2} \vort{} \, ,
\end{equation*}
where the vorticity is:
\begin{equation*}
  \vort{} = \nabla \times \bfg v = 2\omega_o \, \unitvec{\gamma} \, .
\end{equation*}
At this point, it is relevant to observe that the expression of the velocity $\bfg v(t, \bfg \varphi(t))$ is not unique.  In fact, one can also write the voltage as:
\begin{equation*}
  \bfg v_h(t, \bfg \varphi(t)) =
  \begin{bmatrix}
    -h\omega_o \, \varphi_{\beta}(t) + h \zeta_h \, V \cos \theta(t) \\
    h\omega_o \, \varphi_{\alpha}(t) + h \zeta_h \, V \sin \theta(t) \\
    0
  \end{bmatrix} ,
\end{equation*}
which leads to rewrite the vorticity as:
\begin{equation*}
  \vort{}_h = \nabla \times \bfg v_h = 2h \omega_o \, \unitvec{\gamma} \, .
\end{equation*}
While surprising at first, this result is justified by the fact that the vorticity, unlike the azimuthal frequency $\bfg \omega_v$, is not an invariant.  The vorticity depends on the reference frame chosen to describe the motion for the stream line.  For this reason, the vorticity can be $2\omega_o$ or $2h\omega_o$ depending on the coordinates.  In ac circuits, the natural choice for the reference frame is the one that rotates at $\omega_o$, i.e., the fundamental frequency of the system.  Then the time-varying components:
\begin{equation*}
  \begin{aligned}
    \rho_{h}
    &= \frac{\omega_o V (h-1) V_h \sin(\theta_h - \theta)}
    {V^2+ V_h^2 + 2VV_h \cos(\theta_h - \theta)} \, , \\
    \bfg \omega_{h}
    &= \omega_{h} \, \unitvec{\gamma} =
      -\frac{\omega_o (h-1)V_h (V \cos(\theta_h-\theta) + V_h)}
    {V^2+ V_h^2 + 2VV_h \cos(\theta_h - \theta)} \, \unitvec{\gamma} \, ,
  \end{aligned}
\end{equation*}
can be interpreted as measures of the harmonic distortion with respect to a balanced fundamental frequency.  In particular, the first term defines the distortion of the magnitude and the second term the distortion of the frequency.  Note that one can calculate the distortion for each harmonic individually as well as the total distortion for systems with multiple harmonics, as follows:
\begin{equation*}
  \begin{aligned}
    \rho_t = \sum_{h\in\mathcal{H}} \rho_h
    &= \frac{\omega_o V \sum_{h\in\mathcal{H}}^m(h-1)V_h \sin(\theta_h - \theta)}
      {|\bfg v|^2} \, , \\
    \omega_t = \sum_{h\in\mathcal{H}} \omega_h
    &= -\frac{\omega_o \sum_{h\in\mathcal{H}}^m(h-1) V_h(V \cos(\theta_h-\theta) + V_h)}
    {|\bfg v|^2}\, ,
  \end{aligned}
\end{equation*}
where $\mathcal{H}$ is the set of harmonics present in the voltage.

\begin{figure}[htb]
  \resizebox{0.925\linewidth}{!}{\includegraphics{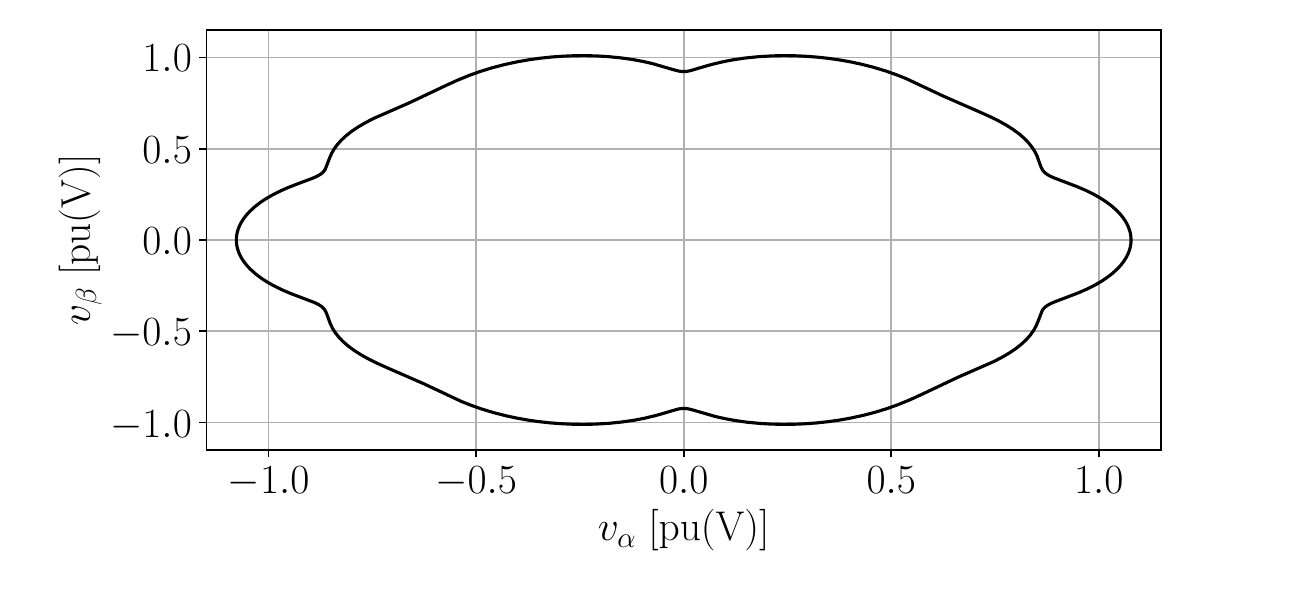}} \\
  \resizebox{0.925\linewidth}{!}{\includegraphics{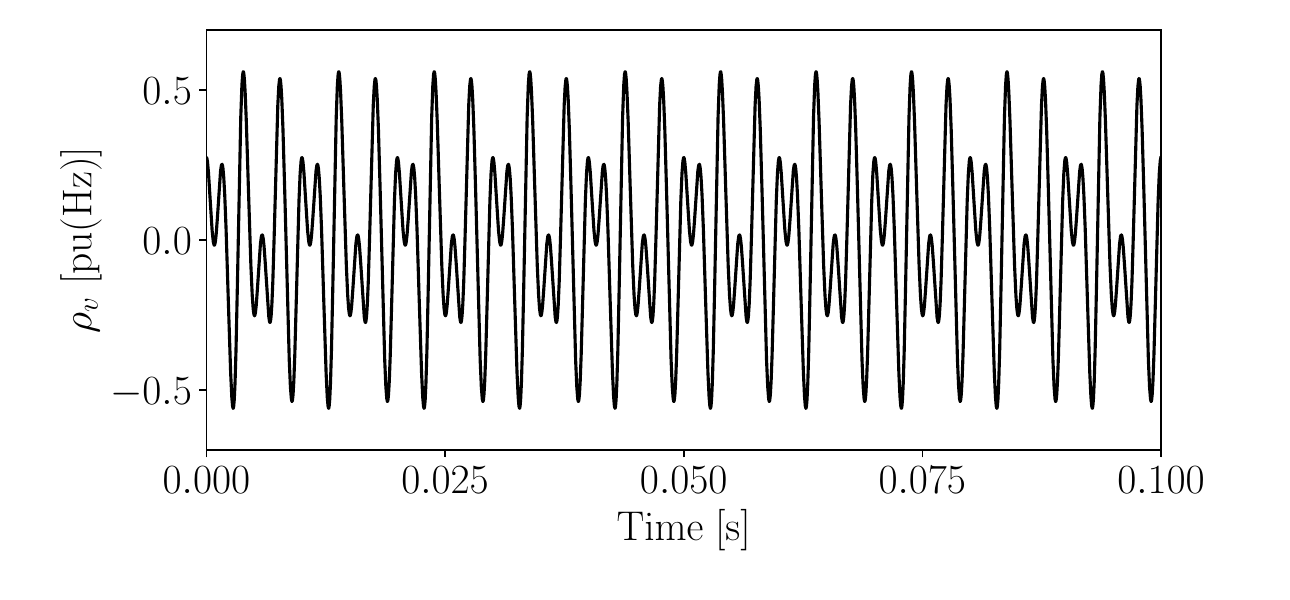}} \\
  \resizebox{0.925\linewidth}{!}{\includegraphics{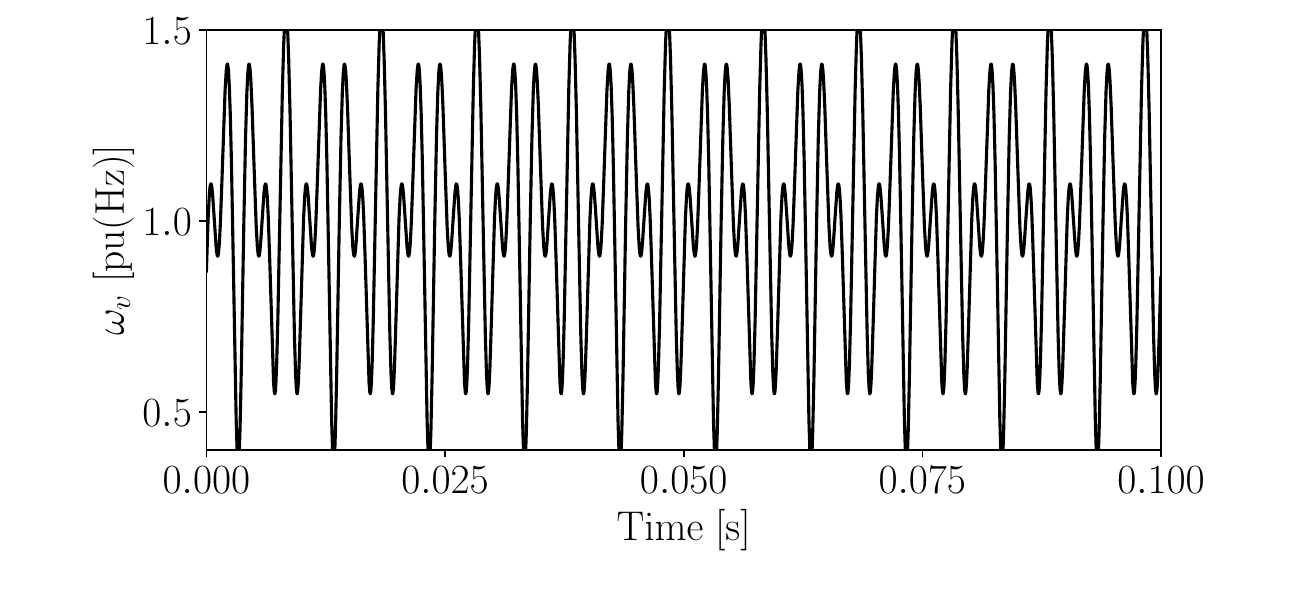}} \\
  \resizebox{0.925\linewidth}{!}{\includegraphics{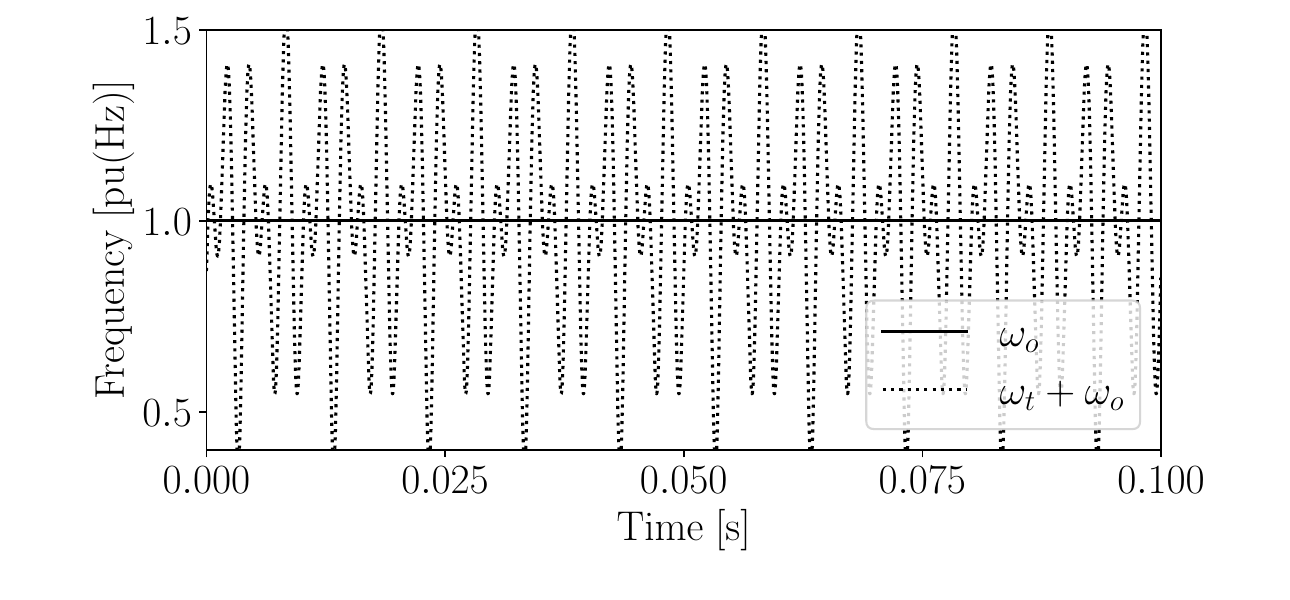}}
  \caption{Stationary balanced non-sinusoidal ac voltage and components of its geometrical frequency and Lagrange derivative.}
  \label{fig:harmonics}
  \vspace{-3mm}
\end{figure}

Figure \ref{fig:harmonics} shows an example of balanced non-sinusoidal case with two harmonics plus the fundamental frequency voltage.  Parameters are: $\mathcal{H} = \{7, 11\}$, $V_h = V/(3h)$, $\omega_o = 2\pi \, 50$ rad/s, $\phi = \pi/6$, $\phi_h = h \phi$.

This example illustrates again the breakdown of the geometric frequency obtained using the Lagrange derivative and fluid mechanics formalism.  The components are as follows.
\begin{enumerate}[(i)]
\item If the reference frame is properly chosen, half the magnitude of the vorticity is the fundamental frequency ($\omega_o$) of the voltage.  The vorticity is constant if the fundamental frequency is constant.
\item $\omega_t$ is the ripple of the frequency due to harmonic content.  This content can be split into each component per harmonic.
\item $\rho_t$ is the radial ripple of the radius of the voltage curve in the $(\alpha, \beta)$ plane.
\end{enumerate}

\vspace{-2mm}

\subsection{Concluding Remarks}

Based on the results of the examples, the following concluding remarks are relevant.

\begin{enumerate}[(i)]
\item All stationary ac voltages can be associated to a solenoidal field or, in the language of fluid mechanics, to the flow of an incompressible fluid.
\item The vorticity is constant in all stationary cases, as opposed to the azimuthal frequency obtained with a purely differential geometry approach.  This is well in accordance with the intuition that in stationary conditions where the voltages have a constant fundamental frequency, the fundamental ``rotation'' is also constant.  This results is also in accordance with a theorem due to Cauchy that proves that the vorticity is the mean value of the rates of rotation of perpendicular axes \cite{Truesdell:1954}.
\item The vorticity is not an invariant, i.e., its expression depends on the choice of the coordinates.  Thus, in general, it is different from $\bfg \omega_v$, which is invariant.  In practical applications, however, it is straightforward to choose a set of coordinates for which the interpretation of the vorticity has a clear physical meaning.
\item In unbalanced conditions, half the magnitude of the vorticity ($\xi \omega_o$, with $\xi >1$) does not coincide with the fundamental frequency.  Rather, the term $\xi \omega_o$ represents an equivalent frequency of a balanced voltage after removing the distortion due to the unbalance.  The scale factor $\xi$ is a function of the unbalance of the magnitude of the voltages.  In common situations, $\xi \approx 1$. 
\item In unbalanced conditions, $\omega_r \ne 0$ and $\rho_r \ne 0$, whereas $\omega_t = 0$ and $\rho_t = 0$.
\item In non-sinusoidal conditions, $\omega_r = 0$ and $\rho_r = 0$, whereas $\omega_t \ne 0$ and $\rho_t \ne 0$.
\item The conditions indicates in the two dual points above allow a straightforward identification of the operating condition of the system.
\end{enumerate}

\section{Conclusions}
\label{sec:conclusions}

This work provides the foundations for a formal equivalence between the geometric frequency of the voltage or current of an electric circuit and the symmetric and skew-symmetric terms that compose the Lagrange derivative of a stream-line of the flow of a fluid.  This equivalence highlights that the radial and azimuthal frequencies that compose the geometric frequency can be further decomposed into terms that have precise physical and/or geometrical meaning.  These are local time-variance $\dt \bfg v$, normal strain (compression/expansion), shear strain (radial and tangential distortion) and rigid-body rotation (vorticity).  These terms are useful to interpret the operation of an electric circuit.  Specifically, if the distortion is not null, the voltage is unbalanced, if $\dt \bfg v \ne \bfg 0$, the voltage has harmonics.  The work also shows that a stationary balanced sinusoidal ac circuit is equivalent to a solenoidal field and to an incompressible fluid, whereas a dc voltage is equivalent to an irrotational field.

We believe that this work constitutes the seed for a novel approach to study ac circuits.  The formal analogy with fluid mechanics provides a well established discipline from which one can obtain theoretical tools as well as analysis techniques.  Future work will delve into this discipline to gain further insights in the study of electrical circuits.  Another work direction that will be pursued is the applications of the proposed approach to practical problems such as power quality improvement and control of electrical devices and networks.

Finally, it is relevant to note that the framework of the Lagrangian derivative requires to write the expressions of the elements of the generalized velocity vector (voltages or currents) as functions of the elements of the generalized position vector (fluxes or charges).  This happens to be a challenging problem to solve, especially when considering voltages and currents for which the generalized positions cannot be measured directly.  A hint on how to solve this problem is given by the Cauchy's theorem mentioned in the remarks of the case study that proves that the vorticity is, in effect, an \textit{average value} of the rate of rotation of the stream-line.  Future work will focus on exploiting this theorem to evaluate the vorticity based on voltage and current measurements.

\appendix

\section{Clarke's Transform}
\label{app:clarke}

The Clarke's transform, often also called $\alpha \beta \gamma$-transform, applied to three-phase voltages is as follows \cite{Clarke:1943}:
\begin{equation}
  \bfg v_{\alpha \beta \gamma }(t) = \bfb C \, \bfg v_{abc}(t) =
  \frac{2}{3}
  \begin{bmatrix}
    1 & -\frac{1}{2} & -\frac {1}{2} \\
    0 & \frac{\sqrt{3}}{2} & -\frac{\sqrt{3}}{2} \\
    \frac{1}{2} & \frac{1}{2} & \frac{1}{2} \\
  \end{bmatrix}
  \begin{bmatrix}
    v_{a}(t)\\
    v_{b}(t)\\
    v_{c}(t)
  \end{bmatrix} ,
\end{equation}
where $\bfg v_{abc}(t)$ is a three-phase voltage vector and $\bfg v_{\alpha \beta \gamma }(t)$ is the corresponding voltage vector given by the transformation $\bfb C$.  The inverse transform is:
\begin{equation}
  \bfg v_{abc}(t) = \bfb C^{-1} \bfg v_{\alpha \beta \gamma }(t) =
  \begin{bmatrix}
    1&0&1\\
    -{\frac {1}{2}}&{\frac {\sqrt {3}}{2}}&1\\
    -{\frac {1}{2}}&-{\frac {\sqrt {3}}{2}}&1
  \end{bmatrix}
  \begin{bmatrix}
    v_{\alpha }(t)\\v_{\beta }(t)\\v_{\gamma }(t)
  \end{bmatrix} .
\end{equation}
This formulation of the Clarke's transform preserves the amplitude of the electrical quantities.  For example, consider the following three-phase balanced voltage: 
\begin{equation}
  \begin{aligned}
    v_{a}(t) &= \sqrt {2} V\cos \theta (t) \, ,\\
    v_{b}(t) &= \sqrt {2} V\cos \left(\theta (t)- \tfrac {2}{3} \pi \right) \, ,\\
    v_{c}(t) &= \sqrt {2} V\cos \left(\theta (t)+ \tfrac {2}{3} \pi \right) \, ,
  \end{aligned}
\end{equation}
where $V$ is the rms magnitude of $v_a(t)$, $v_b(t)$ and $v_c(t)$, and $\theta (t)$ and arbitrary function of time.  In $\alpha \beta \gamma$ coordinates, the voltage becomes:
\begin{equation}
  \begin{aligned}
    v_{\alpha}(t) &= \sqrt{2}V\cos \theta (t) \, ,\\
    v_{\beta}(t) &= \sqrt{2}V\sin \theta (t) \, ,\\
    v_{\gamma} &= 0 \, .
  \end{aligned}
\end{equation}



\begin{IEEEbiography}[{\includegraphics[width=1in, height=1.25in, clip, keepaspectratio]{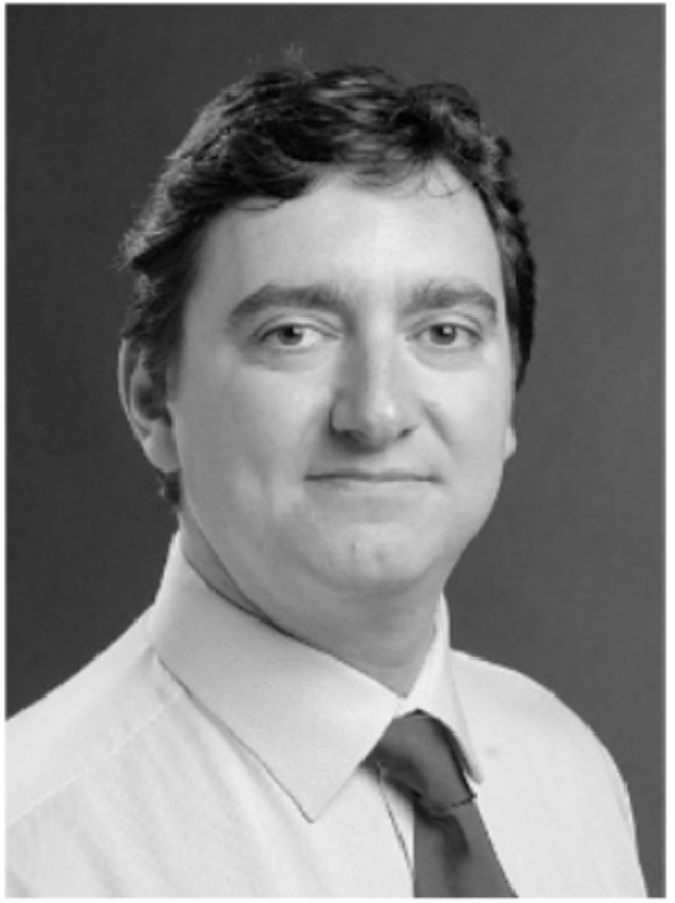}}]
  {Federico Milano} (F'16) received from the University of Genoa, Italy, the ME and Ph.D.~in Electrical Engineering in 1999 and 2003, respectively.  In 2013, he joined the University College Dublin, Ireland, where he is currently a full professor.  He is an IEEE PES Distinguished Lecturer, a senior editor of the IEEE Transactions on Power Systems, an IET Fellow and editor in chief of the IET Generation, Transmission \& Distribution.  He was the chair of the IEEE Power System Stability Controls Subcommittee and of the Technical Programme Committee of the 23th Power System Computation Conference.  His research interests include power system modeling, control and stability analysis.
\end{IEEEbiography}

\vfill

\end{document}